\documentclass{article}

\newcommand{\ket}[1]{{|#1\rangle}}

\begin{document}

\title{Comment on ``Quantum Optimization for Combinatorial Searches''}

\author{Christof Zalka \\
{\tt zalka@uwaterloo.ca} \\
Department of Combinatorics and Optimization, \\
University of Waterloo,
Waterloo, Ontario, Canada N2L 3G1 \\ \\
Todd Brun \\
{\tt tbrun@ias.edu} \\
Institute for Advanced Study, \\
Einstein Drive, Princeton, NJ  08540}


\maketitle

In a recent publication \cite{opt} C.A. Trugenberger claims to have
found a ``quantum optimization'' algorithm which outperforms known
algorithms for minimizing some ``cost function''. Unfortunately, this
algorithm does not work. It is no better than choosing a state at
random and checking whether it has low cost; in fact, carrying out the
procedure can do considerably worse than this.

In the given algorithm, a particular state is prepared on a quantum
computer.  One then makes two measurements in sequence.  Only if
the first measurement results in a desired outcome, do we proceed;
otherwise we start over from the state preparation stage.
The probability of obtaining the desired final state is the product
of the success probability in the first measurement and the probability of 
obtaining the desired state in the second measurement, given success
in the first.  But it is easy to see that for any state, this
overall success probability is {\it at most} one over the number
of possible states.

The author constructs a uniform amplitude superposition of all
$N$ states (his Eq. 8):
\[
\ket{\psi_0} = \frac{1}{\sqrt{N}} \sum_{k=1}^N \ket{I^k}
\]
Upon measuring, we would obtain each state $I^k$ with equal
probability $1/N$. Now the author adds an auxiliary register and
performs a unitary transformation, based on a particular cost function,
which produces a final state of the form (his Eq. 15):
\[
\ket{\psi_{fin}} = \frac{1}{\sqrt{N}} \sum_{k=1}^N \ket{I^k,\phi(I^k)}
\]
He first measures the second (auxiliary) register.  If this is found
to be $\ket{00\cdots0}$, he then measures the first register, obtaining
a state $I^k$, which he hopes to be the state $I^{\rm min}$ which
minimizes the cost function.  More generally, he wants to find a state
$I^k$ which is {\it close} to optimal, i.e., which has cost
$C(I^k)<C^{\rm tol}$ for some low value $C^{\rm tol}$.  Let's suppose
there are $M$ such states.

The two-step procedure, however, is unnecessary---it could
just as well be done in one go by measuring both registers
simultaneously, and possibly rejecting the result depending
on what is found in the second register.  The probability of one
attempt being successful is then
\begin{eqnarray}
p(C(I^k)<C^{\rm tol}|0\cdots0) \cdot p(0\cdots 0)
  &=& p(C(I^k)<C^{\rm tol} \& 0\cdots0) \nonumber\\
  &=& p(C(I^k)<C^{\rm tol}) \cdot p(0\cdots 0 | C(I^k)<C^{\rm tol}) \nonumber\\
  &=& (M/N) \cdot p(0\cdots 0| C(I^k)<C^{\rm tol}) \nonumber\\
&\le& M/N \;,\nonumber
\end{eqnarray}
(where $p(A\&B)$ denotes the probability of outcomes $A$ and $B$ both
occurring, and $p(A|B)$ the probability of $A$ given that $B$ has occurred).
The probability of a successful outcome is less than or equal
to $M/N$, which is just the probability of measuring the
first register and randomly getting an $I^k$ with $C(I^k)<C^{\rm tol}$.
The algorithm is therefore worse than a random search---possibly much worse.

In the paper, the success probabilities for the two measurements to yield any
fixed state $I^k$ are given by $P^0_b$ (Eq. 16) and $P_b (I^k)$ (Eq. 17), and
clearly their product is less than (or equal to) $1/N$.

Finally we would like to point out that any algorithm with an
interesting performance for minimizing a cost function, has to
exploit some structure of this function. Heuristics typically rely on
the tendency of ``neighboring'' states to have similar cost. But the
author doesn't exploit anything like that. He considers the states
as a set without any structure, in particular without a metric.


\end{document}